# Computer Aided Detection and Classification of mammograms using Convolutional Neural Network

Kashif Ishaq; Muhammad Mustagis

**ABSTRACT** Breast cancer is one of the most major causes of death among women, after lung cancer. Breast cancer detection advancements can increase the survival rate of patients through earlier detection. Breast cancer that can be detected by using mammographic imaging is now considered crucial step for computer aided systems. Researchers have explained many techniques for the automatic detection of initial tumors. The early breast cancer symptoms include masses and micro-calcifications. Because there is the variation in the tumor shape, size and position it is difficult to extract abnormal region from normal tissues. So, machine learning can help medical professionals make more accurate diagnoses of the disease whereas deep learning or neural networks are one of the methods that can be used to distinguish regular and irregular breast identification. In this study the extraction method for the classification of breast masses as normal and abnormal we have used is convolutional neural network (CNN) on mammograms. DDSM dataset has been used in which nearly 460 images are of normal and 920 of abnormal breasts.

## I. INTRODUCTION

Cancer is the world's most prevalent cause of morality among humans. Cancer can affect various parts of human body in which breast cancer is the second major cause of death among women. The likelihood of survival is diminished if diagnosis is not made in the early stages. [1]. In females the prevalent cause of mortality is the breast cancer. There may be chances to control cancer at an earlier stage through early treatment of this disease which is only possible if its detected in early stages. [2]. According to survey across Pakistan almost every year there are 80% new cases found and majority of these cases can be remedied through early identification which can increase the life expectancy up to 90% [1][3]. Medical images are most commonly studied images and these images are used to study and extract desired results to detect and cure various disease. We use them either for extraction of features or to conduct certain morphological surgery on pictures. Breast tumor is very difficult to detect and handle because their diagnosis varies with one another. Many software methods have been used to detect and cure breast cancer. These methods include [3]: thermal imaging, ultrasound imaging, surgical biopsy, magnetic resonance imaging, mammography X-ray imaging and computed tomography imaging. CAD is largely used to diagnose cancer at its initial stages. Each of these methods have several limitations to achieve the desired results [4][7]. The figure displays breast cancer images that were created using various imaging techniques.

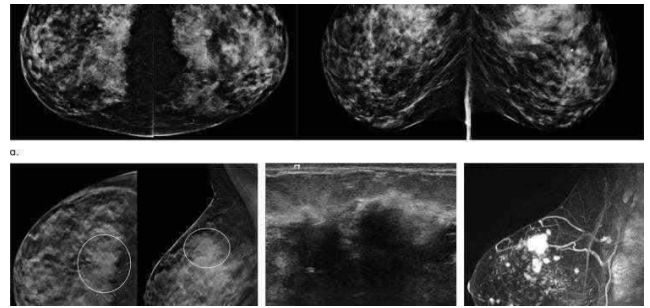

**FIGURE 1.** *Tumor classification using various imaging Techniques*

Mammograms breast diagnosis is most efficient for gathering information through automated and balanced examination of any breast tissue aberration in medical image processing [5][6]. To understand the impact of the patient's anatomy following correct detection, it is necessary to comprehend tumor conditions in image diagnosis [3][5]. The various factor that limits the proper detection of the disease through mammograms. So, to overcome this problem we have come with the computer aided detection systems [6]. Computer aided detection system will dissect all the mammograms to determine the presence or absence of a cancerous tumor in the area being examined. The software will highlight the potential areas of interest. The radiologist will find two major i-e Masses and micro-calcification signatures. The small amount of calcium deposits in mammograms also termed as micro-calcification is one of the key identifiers for spotting cancer at its earliest stage [8]. The technique used for extraction of features provide mathematical characteristics, textural and graphical data. [5]. The various forms of classifications such as Naïve Bayes probabilistic classifier, SVM, decision trees and K-Nearest relative techniques. Both such classifiers were used and obtained the desired outcomes [9][10]. Deep Learning (DL) model provides such platform that learn and reuse the data to train models without human intervention to obtain desired results [11]-[13]. CNN is majorly used for image classification [15]. Some of the variants of CNNs are LeNet, VGG, GoogleNet, AlexNet and ResNet, which pre-trained on ImageNet dataset [14][16]. Deep neural network (DNN) can self-learn the features and perform task based on the dataset provided to achieve desired outcome [17]. Neural networks have replaced previous techniques to bring advancement in this area [18]. CNN contain large number layers based on huge dataset provided and each layer extract data to detect cancerous cell.

The next section of this paper is formulated as: Section II in this section the research objectives, methods and reason why the research has been carried out has been discussed. The part discusses the research questions, procedure to select the relevant articles for research based on relative keyword selection. In section III experimental methods and results has been discussed to analyze the research problem systematically. In section IV the comparative analysis of various methods to the proposed method, future frame work has been discussed. In section V the review has been concluded.

## II. RESEARCH METHODOLOGY

In this study we will be working on X-ray imaging procedure called breast photos Mammograms. Our data collection contains nine-hundred (900) images of the tumor and the non-tumor. Our final product would use Google Collaboratory platform to simplify the breast tumor detection by utilizing Deep Neural Networks. For less drawbacks it can offer the best performance. Images of breast mammograms include three groups of I Regular, Benign, and Malignant. We can use the CNN method to identify a group of Automated Breast Tumors.

### A. RESEARCH OBJECTIVES (RO)

The main objectives of this research are as follows:

RO1: The main thought behind this study is to develop a technique based on mammographic Imaging for the detection and classification of breast cancer using Deep Neural Networks.

RO2: Comparative analysis of the related research that has been done in related area of interest and CNN classifiers being used in these researches.

RO3: Illustrate how this research can overcome the prevalent data, as well as offer a way in rendering network forecasts more explicable

RO4: The scope and challenges that must be catered through future work in the area of interest.

### B. RESEARCH QUESTIONS (RQ)

The research questions have been addressed to establish the prime motivation behind this study. These questions help understand the method adopted for the study more effectively. Research questions are given in below table.

**TABLE I.** RQ and major motivations

| | Research Question | Motivation |
|---|---|---|
| RQ1 | What were the prime data sources or publication channels used in the area of concern to validate the authenticity of the research? | To understand the importance of research that has been done and why further increments are required in the relevant area. |
| RQ2 | What are the research techniques used previously to study mammography and what techniques are used during current research study? | To understand the domain of the research and the prevalent study done to minimize the research gap. |
| RQ3 | How the quality assessment criteria were imposed on the research? | To develop an understanding of work and effort that has been put in this area under observation |
| RQ4 | How the current research work adds to previous researches and comparative analysis of research with previous methods? | To understand the techniques used previously and how our research aids the area of study |

### C. SEARCH SCHEME

The search investigation for the required area of interest has been carried out by focusing on the prime factors/ keywords that impact our research. These keywords help us out to filter the right set of data focusing on concerned studies that must be engaged to conduct the study. To get optimal results the research keywords were used on Web of sciences database to conduct a systematic review. The screening on the record to extract out the data that is considered was done through proper selection of keywords. The table II enlist the terms used and the synonyms/ alternate terms that were focused are shown below.

**TABLE II.** keywords used as strings

| Keyword | Alternate Keywords |
|---|---|
| Computer-aided diagnosis | Features segmentation/ classification, feature extraction |
| Mammography | Mastography, Digitized mammographic images |
| Convolution neural network | Deep convolution neural network, deep neural networks (DNN), Deep Learning |
| Breast Cancer | - |

The strings that were finalized were used to find the related publications and studies that were carried out while also studying the extracted keywords to confine the desired results related to computer aided detection, convolution neural networks and results regarding mammography were studied.

1) ASSESSMENT OF RESEARCH QUESTION 1. WHAT WERE THE PRIME DATA SOURCES OR PUBLICATION CHANNELS USED IN THE AREA OF CONCERN TO VALIDATE THE AUTHENTICITY OF THE RESEARCH?

The below table enlist the publication channels and the source of data used for the study. For the research various journals, articles, conference papers were studied. It also includes data from various online accessed websites that were helpful during the study.

TABLE III. Publisher Sources used during research

| Sr No | Publication Source | No of Publications |
|---|---|---|
| 1 | Journal of Biomedical Informatics | 1 |
| 2 | IEEE J Biomed Health Inform | 2 |
| 3 | BMC Cancer Journal | 1 |
| 4 | Oriental journal of computer science and technology | 1 |
| 5 | Computational Intelligence and Neuroscience journal-NCBI | 5 |
| 6 | in International Conference on Emerging Trends in Information Technology and Engineering | 2 |
| 7 | Journal of Medical Image Analysis | 1 |
| 8 | Journals of nature | 3 |
| 9 | Journal of algorithms | 1 |
| 10 | Journal of Procedia Computer Science | 1 |
| 11 | Journal of Neural Networks | 1 |
| 12 | Biomedical & Pharmacology Journal | 1 |
| 13 | magnetic resonance imaging journal | 1 |
| 14 | International Journal of Computer and Electrical Engineering | 2 |
| 15 | IJISAE | 1 |
| 16 | Journal of Soft computing | 1 |
| 17 | IEEE TRANSACTIONS ON NANOBIOSCIENCE | 1 |
| 18 | Computer Methods and Programs in Biomedicine | 1 |
| 19 | International journal of medical informatics | 2 |
| 20 | IEEE Access | 2 |
| 21 | Biomedical Signal Processing and Control | 1 |
| 22 | International Journal of Multimedia and Ubiquitous Engineering | 1 |
| 23 | Multidisciplinary Cancer Investigation, | 1 |
| 24 | Academic Radiology | 1 |
| 25 | ICT Express | 1 |
| 26 | Expert Systems with Applications | 1 |
| 27 | IEEE transactions on Instrumentation and Measurement | 1 |
| 28 | Annals of biomedical engineering | 1 |
| 29 | Journal of Electronic Imaging | 1 |
| 30 | Computational and mathematical methods in medicine | 1 |
| 31 | Information Fusion | 1 |
| 32 | Journal of digital imaging, | 1 |
| 33 | Research on Biomedical Engineering | 1 |
| 34 | Informatics in Medicine Unlocked | 1 |
| 35 | Computer Methods and Programs in Biomedicine | 1 |
| 36 | Expert Systems with Applications | 1 |
| 37 | IEEE Transactions on Image Processing | 1 |
| 38 | ACM Computing Surveys | 1 |
| 25 | Conference | 10 |
| 26 | Accessed online | 19 |

2) ASSESSMENT OF RESEARCH QUESTION 2. WHAT ARE THE RESEARCH TECHNIQUES USED PREVIOUSLY TO STUDY MAMMOGRAPHY AND WHAT TECHNIQUES ARE USED DURING CURRENT RESEARCH STUDY?

A lot of research has been done to fill the research gap through various techniques and deep learning models for the classification of mammograms through feature extraction and feature segmentation. The data set obtained from images are put under study by using CNN model to distinguish abnormal images from normal ones.

The study conducted to reduce the gap between mammogram identification and segmentation using wavelet neural network approach. The study was conducted on 216 clinical data images of mammograms. The algorithm works by focusing on the region of interest and leading to distinguishing normal mammograms from abnormal ones [19][20].

The research conducted on automated segmentation of the breast tumor from the mammogram photographs using modern fuzzy algorithm. This algorithm is implemented to determine the inconsistencies in image along tumor field boundaries. The segmentation is done by the values obtained in the fuzzy region. The values obtained distinguish on the bases of values obtained for a healthier tissue as compared to affected one [21].

Automated breast cancer classification is improved by expanded white matter segmentation; gray matter lesion segmentation is provided on mammogram data collected by k-nearest neighbor classifier. Classification is achieved by preparation is performed by mammogram diagram classification to emphasis. Faults in classification were not observed in 98% of breast tumor classification and 97% of white matter classification [22].

In another research, neural network is use to design a prototype of tumor and non-tumor area. The proposed research classifies tumor as benign and malignant through pattern- recognition. It classifies best by finding optimal activation functions that reduce the classification error. Then results are improved by using CMOS technology. They progress in separate approaches as gradient vector flow and boundary vector flow, get certain image concavities and forms and then evaluate the image vector flow [23].

The Fuzzy C-implies model is used to attempt an innovative breast tumor segmentation process. To compare the effects of the proposed procedure with other FCM approaches, they utilize simulated and actual breast pictures, mammograms, and monitor noise rates. Simulated and actual breast pictures, mammograms, and noise rates are used to compare the effects of the proposed procedure with other FCM approaches [24].

In another study conducted on the color segmentation technique for the classification of mammograms. The grayscale mammogram images were converted in to RBG images. These colored images were then equated in to high and lower values using histogram clustering. Through this image segmentation the position of the tumor was identified [25].

A hybrid technique termed as k-SVM which utilizes k means and support vector machines to segment between abnormalities found in the breast area by k-mean clustering. The breast lesion is segmented in to four clusters to identify tumor [26].

In this study a machine learning technique was used to distinguish between a normal and tumor affected mammogram images though image digitizing, pre-processing and segmentation [27].

A study conducted in 2016 uses neural network techniques to train the data to identify cancerous images through pattern recognition. The image vector flow was evaluated and the results obtained were improved using CMOS technology [28].

A study published in soft computing journal proposed a segmentation technique for the breast cancer identification. The pattern recognition and Fuzzy C mean algorithm. The algorithm treated input data to train and test the model by grouping images in to two instances (benign and malign). The proposed methodology showed high accuracy in breast classification [29].

The study conducted by researchers used top hat transformation and gray level co-occurrence matrix to categorize the tumor, and compared their results with back propagation neural networks. The use of observe classes has enabled the detection of the tumor, which is highly dependent on future extraction methods. The research utilizes two data sets, DDSM and MIAS. By presenting a robust classification model, the objective is to reduce false assumptions in medical informatics. 91.5% accuracy was achieved by the research in determining whether it was benign or malignant [30].

A transfer learning approach was used for classification. In transfer learning the model reuses the data to train the model. The hybrid approach by training the model on spine image data using CNN and then using this trained model to classify between mammogram images. This extracted data is then used to identify tumor by using Support vector machine with radical basis function (RBF) [31].

Various methods are used in this paper to classify breast tumors. By using the histogram equalization technique, they enhance the mammogram and suppress noise. Feature extraction methods using discrete cosine transform (DCT) are utilized in the proposed research. The abnormal tissue in the tumor area is classified as either benign or malignant using a Bayesian classifier. The research is conducted based on the MIAS image data set. The system provides efficient results, as evidenced by the results [48].

The research has been done to improve the image quality using Deep Neural Network with Support Value (DNNS) stating DNNS is efficient than any other existing method. Data augmentation technique has been used to enhance the prediction of hispathalogical images. The research was conducted on 638 samples of hispathalogical images. Through these enhanced images, the textural and geometrical feature are extracted using extraction techniques. The classification is done by using support value of deep neural network [53].

In this paper, for the classification of benign and malignant tumors artificial neural networks uses single and multi-layer perceptron networks. First Figure 1. Tumor classification using various imaging Techniques the pixels are used to calculate boundary shape of the effected region. The textural information is withdrawn through GLCM. Each neural network layer sorts mammogram non-cancerous and cancerous image of the organ. The results are extracted using k-nearest neighboring technique for its segmentation [59].

The comparison of the various studies is shown in the table below. the table compares the pre-processing and classification techniques used and the results that were achieved through these methods. It also discusses the area of focus during that research and how these results were relative to the study conducted during this research.

TABLE III. **Comparative study that has been conducted previously**

| Sr No # | Title | Focus of research | Pre-Processing Techniques | Classification | Accuracy /Results |
|---|---|---|---|---|---|

| # | | | | Techniques | |
|---|---|---|---|---|---|
| 1 | Neural network And multi-fractal dimension features for breast cancer classification from ultrasound images | Tumor Classification | >Median filter Removing any incautious artefacts from the area >Adaptive weighted median filter | ANN classifier | The classifier yielded a high precision of 82.04%, Sensitivity of 79.3991%, and specificity of 84.7559% |
| 2 | Deep learning-based computer aided diagnosis system for breast mammograms | Cancer Detection | Enhancement process Preserves the edges by using Bilateral filter information by resized to suitable size 96x96 using a special technique bicubic interpolation | Deep CNN has been used with SVM for features extraction and classification | MIASmini + DDMS = 93.35 |
| 3 | A Region Based Convolutional Network for Tumor Detection and Classification in Breast Mammography | Early Detection of Breast Cancer | >Median filtering to reduce "salt and pepper" noise. >Image interpolation techniques to enhance the quality of digital images | Neural Networks, J48 Decision Trees, Random Forest and K- Nearest Neighbor | >Neural Networks based Classification Accuracy= %96. >48-Decision Trees based Classification Accuracy = 95.7055% |
| 4 | Automated Breast Ultrasound Lesions Detection Using Convolutional Neural Networks | Lesions Detection | Noise was partly reduced by the US acquisition system Breast ultrasound images are in grayscale and are split into 28×28 patches | CNN-Layers >Patch-based approach using LeNet >transfer learning FCN-AlexNet | LeNet: Dataset A(TFP)= 0.89, (FP)= 0.10 FCN-AlexNet: Dataset A (TFP)= 0.98, (FP)= 0.16 |
| 5 | Breast Cancer Detection using Image Processing Techniques | Cancer Detection | converted into grayscale image of 2D matrix Noise removal algorithm | >Neural Network (NN) >Back Propagation Neural Network | DDMS = 93.35 |

## III. MATERIAL AND METHODS

### D. DATA SETS

The mammographic imaging society (MIAS) image data set consist of 330 scans mammogram images containing normal, benign and malignant classes. [72] These scans are set to the size of 1024x1024 pixels. Due to the small amount of data set, it is very difficult to use it for training. Therefore, we use DDSM data set. It consists of 2620 scanned mammography images. Information collection includes several normal, benign and malignant images of each class. The data set contains full medical information of each image. The data set can also be download as GIF images.[73] The images of DDSM also contain the subset of mammographic imaging society (MIAS) images. These images are expanded and can be save in DICOM format. These images contain the region of interest (ROI) already extracted from each image and can be used for training purposes.

The mammogram images are always taken in these two views that is 'Cranial-Caudal' (CC) is a view, it is taken from above side of breast and the second one 'medio-lateral' view (MLO); an oblique angle view that is taken from the mid of chest. So DDSM data set contain each class image with various patients IDs that is it contains the information of one patient as Right CC view, Left CC view, Right MLO view and Left MLO view. [74]

TABLE IV. Details of DDSM Data Set

| Data Set | Illustrations | Class | Source | Type/Format |
|---|---|---|---|---|
| DDSM | 460 | Normal | TCIA | .GIF |
| DDSM | 460 | Benign | TCIA | .GIF |
| DDSM | 460 | Malignant | TCIA | .GIF |
| | Total = 1380 | | | |

## E. PROPOSED METHODOLOGY

On mammograms we conducted breast photos of separate 'Cranial-Caudal' (CC) and 'medio-lateral' (MLO) perspectives. Production based on breast tumor classification and its normal, benign, and malignant grades.

Our suggested solution achieves the same outcomes with less constraints. Breast tumor data collection has two views to diagnose tumor and after adding the views with separate layers of the neural network we get better results. We use deep neural networks to detect the tumor [75].

General design flow begins from gathering various stages, stopping at the tumor result type.

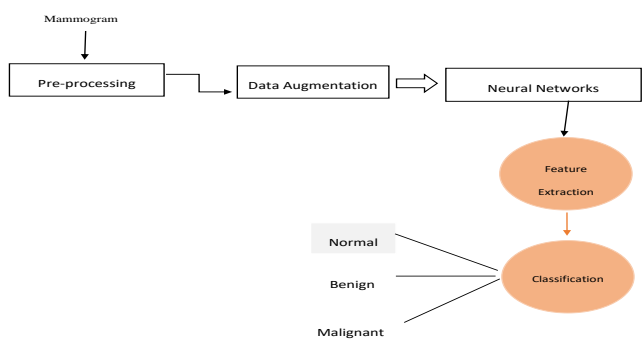

**FIGURE 2.** Flow chart for the detection of breast tumor

### E. 1 CLASSIFICATION OF BREAST TUMOR

Classification of the breast tumors involves various forms and specific architectures. This is why manual classifications are dangerous as doctors plan to recommend medication and more care after all the tests depending on the particular protocol and during this extensive phase. Therefor we move from manual classification to automatic classification. To detect the tumor, biopsy is an important task to detect the disease. The chances of error rate is very high when the radiologist considered the disease manually by just looking from the X-ray images (mammograms) to detect the type of tumor present. Most of the time, they didn't recognize the type of tumor present [76]. Therefore, computer aided detection proven to solve all these kinds of manual issues. In this research we are going to provide automatic computer aided detection and classification of breast tumor by using deep neural networks. Our research uses Convolutional Neural Network layers to classify the type of tumor present in the breast that is either it is normal, benign and malignant. Both tests for an automatic identification of tumors are performed in the Google Collaboratory Method.

Each of the breast images is comparison checked. The breast tumor in mammograms normally appears in a compact shape. The normal mammograms means that there is no risk of any abnormality present in breast tissue. The benign images contain in the mil ducts contain some white small calcifications that does not cross the boundary of breast. The benign size of tumor is 2cm or less than 2cm. This tumor may present in hard or irregular in structure. Benign masses generally possess round, well-circumscribed boundaries and smooth. The malignant tumor is more than >2cm and tumor has spread outside the walls of chest. The malignant tumor, usually have rough, blurry boundaries and spiculated. Therefore, the main focus of research is detecting these three classes by automatic classification. Mammograms contain any abnormality appears as white mass in breast. This white mass called micro calcifications can be a tumor or non-tumor [77]. Now to detect this white mass is benign or malignant, we designed a model there the classification has been done automatically using convolution neural network layers

### E.2 Pre-Processing

Before training CNN, mammogram preprocessing is an important task. It includes removal of unwanted background and enhancing its contrast. As these mammograms already contain the segmented images containing the background area removal and tags. Image enhancement usually done using well known filters like median and histogram Equalization techniques [78].

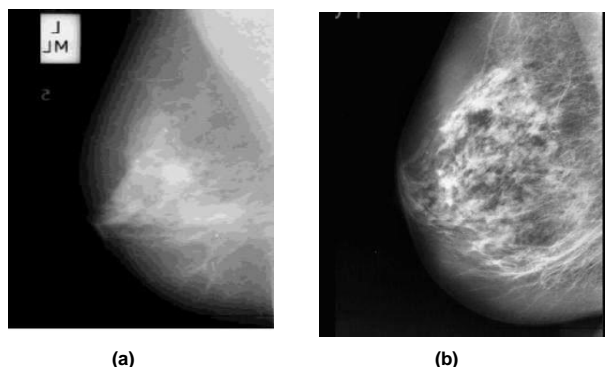

   (a)              (b)
**FIGURE 3:** (a) Original mammogram; (b) Contrasted Mammogram

Many experiments have used segmented ROIs to reduce the CNN's estimation to prevent the problem of minor training results. Such ROIs may be achieved using the available ground truth data, or an automated detection method, through manual segmentation of the images [79].For the lesion concentrated inside the image, the ROIs are clipped and re-scaled to r *r pixels.

Two methods are going to use maximum picture size to train CNNs on mammograms, rather than ROIs. The first technique, high-resolution pictures down-sample to size of 225*225. It is doubtful that the criterion for mammograms to identify limited mass regions or masses in images in down sampling will going to be successful [80]. The second approach is to train a CNN patch-level classifier, which can then be used as an image-level model attribute extractor. The image is partitioned into a series of patches with a slight overlap in the layer-level layout, so each patch is stored completely inside the layer. Full grouping includes sorting through patches and views of the CC & MLO.

3) ASSESSMENT OF RESEARCH QUESTION 3. HOW THE QUALITY ASSESSMENT CRITERIA WAS IMPOSED ON THE RESEARCH?

The quality criteria on which the research was accessed includes various parameters. The research channels used for study included publications in journal and conferences, it also included data from websites that were accessed online. The below table V defines what internal assessment criteria were used and how scoring was done to rate the research channel. The research was rated on the basis of research technique used to achieve the desired results. If the technique used was successful in achieving what it claims the scoring given to it was 1, similarly if the research was able to yield only intermediate results the criteria rated it as 0.5 and if the results not supported the research the rating given to it was termed to 0. Similarly other factors like the empirical type of the research, whether it was qualitative or quantitative based on the study conducted or both approaches were used for the research. The methodology used in the certain publication channel was also rated by analyzing it on basis of the type of method used to achieve results. The data studied mostly varied between few factors like experimental, observational and ground theory.

The pie chart explains the methodology adopted during the research of the articles that were studied. The analysis show that most of the paper that were incorporated for study focused on the experimental methods which was helpful during the research.

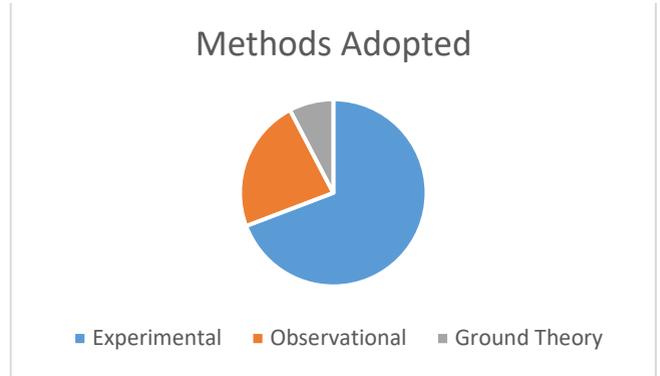

**FIGURE V:** Methods adopted across various research paper that were studied

4) ASSESSMENT OF RESEARCH QUESTION 4. HOW THE CURRENT RESEARCH WORK ADDS TO PREVIOUS RESEARCHES AND COMPARATIVE ANALYSIS OF RESEARCH WITH PREVIOUS METHODS?

In this section of the research the experimentation and the results were discussed to examine research cover any gap over the study that has been previously conducted.

*F. EXPERIMENTATION*
The collection of data provided below referred to as mammograms for breast. The data collection is comprised of three types of regular, healthy, and malignant mammograms. In comparison, all the breast pictures are screened. These scans are set to the size of 1024x1024 pixels. Due to the small amount of data set, it is very difficult to use it for training.
It consists of 2620 scanned mammography images. Information collection includes several normal, benign and malignant images of each class. The data set contains full medical information of each image.

TABLE VI. Data set explanation

| Data Set | Description |
|---|---|
| Normal Images | The normal mammograms contain no white matter present in the images, so there is no sign of tumor in it. |
| Benign Images | It contains different views of mammogram and these images contain the white matter with size less than 2cm but this tumor dose not cross the walls of chest. |
| Malignant images | It also contains the mammograms of all views. These images have white matter that cross the walls of chest with size more than 2cm |

Google Collaboratory allows us to implement our code in python by opening new notebooks to write and execute code. It is a popular cloud software, supporting popular GPU. It is a good compensation when you have no apace in your drives for holding large data sets. The toolkits used in mammography training of CNNs are TensorFlow and Keras. TensorFlow is one of the most open-source Deep learning libraries. TensorFlow library based on Python, able to operate on several CPUs. This may be used directly to build deep learning models using libraries (e.g. Keras). Keras is an open-source library which is used to create powerful deep neural networks and use TensorFlow as the back-end.

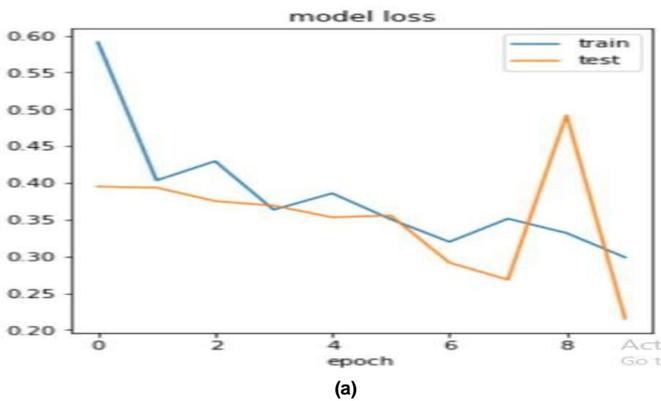

(a)

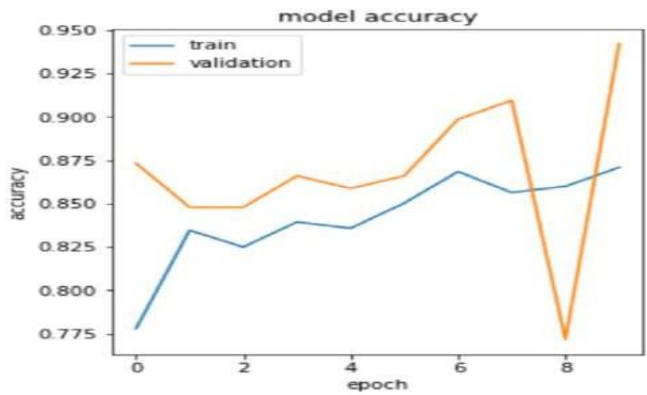

(b)

**FIGURE VI:** (a) Model accuracy; (b) Model loss

*F.1 Experimentation Method*

The proposed model uses Convolutional Neural Network layers for its implementation. The model uses hidden layers of CNN that include 4 convolutional layers, 4, max pooling layers, dropout, flatten and dense layers. These layers are implemented using Google collaborator tool that s implemented in python.

Defining one or more callbacks is helpful before developing the pattern. Model Checkpoint is helpful application. In Model Checkpoint a lot of iterations are always needed because testing requires a lot of time to get a successful outcome. In this scenario, it is safer to just save a copy of the best performing model when an epoch which improves metrics ends.

The model layer details with output of our proposed research are shown below using Adam as optimizer in figure 4.

*F.2 Performance model*

The most important criterion for model output evaluation is accuracy. Our model consists of equal number of mammograms for each class having total of 960 images. We divide our data into 70:30 ratio having 70% on training data and 30% in testing and validation. We trained our model in 10epocs.

To get a clearer understanding of true positives (TP), true negatives (TN), fake positives (FP), and false negatives (FN) for a closer look at misclassification.

Precision is the percentage of accurately expected positive observations to the overall positive observations forecasted.

Precision = True Positive/ (True Positive +False Positive)

Recall is the percentage of accurately expected positives to all real class findings.

Recall = True Positive/ (True Positive +False Negative)

F1-Score is the average of precision and Recall.

F1-Score = (2*Recall*Prescion)/ (Recall + Prescion)

*F.3 Confusion Matrix*

Confusion Matrix is an incredibly significant tool when evaluating misclassification. In this row of the matrix represents the instances in the predicted class, while each column represents the instances inside the class. The diagonals represent groups that were correctly categorized.

*G. RESULTS*

Classification of the mammogram images and the values obtained through the model implemented are listed in the table below

TABLE VII. Results of our proposed model

| Class | Precision | Recall | F1 Score |
|---|---|---|---|
| Normal | 0.91 | 0.89 | 0.89 |
| Benign | 0.94 | 0.90 | 0.91 |
| Malignant | 1.00 | 0.87 | 0.94 |
| Average | 0.95 | 0.88 | 0.91 |

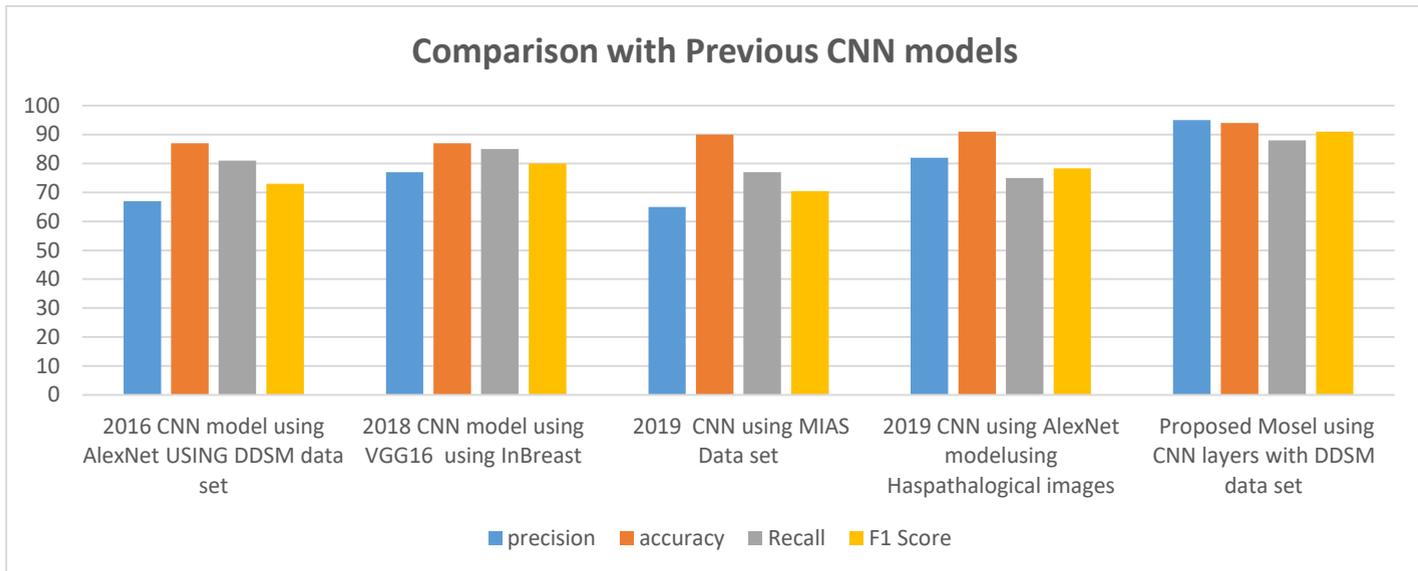

**Chart I:** Comparison with previous CNN models

### H. CONCLUSION

This research, we propose a convolutional neural network for the classification of mammograms as normal, benign and malignant. The experiment is done on DDSM data set tak8ng both views that is CC and MLO. Different sizes of filters can be use in hidden layers of CNN. The work was done on Google Collaboratory python framework. We illustrate how diligent data augmentation and deep neural networks can overcome the prevalent data bottleneck in medical computer vision activities, as well as offer a way in rendering network forecasts more explicable. Our methodology delivers state-of-the-art performance, outclassing professional radiologists, and making for more relaxed acceptance in real-world environments with usability. The efficiency of our model is confirming by seeing the results giving the accuracy of 94% which is significantly good.

### I. FUTURE WORK

Other networks which include the very deep-convolution network (VGG) and the residual architecture (ResNet) will be suggested for future research. Future research involves testing different structures and incorporating processes of focus that are tougher to train but may have much more precise interpretability. tremendous comprehension has been developed about the contribution of the research in the domain. Ultimately, these keywords have been used to classify the articles for review mapping.

```
Edit   View   Insert   Runtime   Tools   Help   Last edited on August 2

de   + Text

Model: "sequential_9"
_________________________________________________________________
Layer (type)                   Output Shape              Param #
=================================================================
conv2d_20 (Conv2D)             (None, 223, 223, 16)      448
_________________________________________________________________
max_pooling2d_16 (MaxPooling   (None, 74, 74, 16)        0
_________________________________________________________________
conv2d_21 (Conv2D)             (None, 72, 72, 32)        4640
_________________________________________________________________
max_pooling2d_17 (MaxPooling   (None, 36, 36, 32)        0
_________________________________________________________________
conv2d_22 (Conv2D)             (None, 35, 35, 64)        8256
_________________________________________________________________
zero_padding2d_7 (ZeroPaddin   (None, 39, 39, 64)        0
_________________________________________________________________
max_pooling2d_18 (MaxPooling   (None, 13, 13, 64)        0
_________________________________________________________________
conv2d_23 (Conv2D)             (None, 12, 12, 64)        16448
_________________________________________________________________
zero_padding2d_8 (ZeroPaddin   (None, 16, 16, 64)        0
_________________________________________________________________
max_pooling2d_19 (MaxPooling   (None, 5, 5, 64)          0
_________________________________________________________________
flatten_4 (Flatten)            (None, 1600)              0
_________________________________________________________________
dropout_4 (Dropout)            (None, 1600)              0
_________________________________________________________________
dense_13 (Dense)               (None, 256)               409856
_________________________________________________________________
dense_14 (Dense)               (None, 128)               32896
_________________________________________________________________
dense_15 (Dense)               (None, 64)                8256
_________________________________________________________________
dense_16 (Dense)               (None, 3)                 195
=================================================================
Total params: 480,995
Trainable params: 480,995
Non-trainable params: 0
```

**FIGURE IV**. Output shape and parameters of our CNN model

TABLE IV. Quality assessment of reviewed article

| | | Internal Assessment | | | | | | |
|---|---|---|---|---|---|---|---|---|
| **References** | **P. Channel** | **Research Technique (a)** | **Empirical Type (b)** | **Methodology (c)** | **(a)** | **(b)** | **(c)** | **Score** |
| (J. Dheeba et al,2014) | Journal | Computer Aided Diagnosis | Qualitative | Experimental | 1 | 0.5 | 0.5 | 2 |
| (M.H. Yap et al., 2019) | Journal | Transfer learning, U-Net, Patch-Based LeNet | Mix Method | Experimental | 0.5 | 1 | 1 | 2.5 |
| (P Giri & K Saravanakumar,2017) | Journal | Computer aided detection using transfer learning | Qualitative | Experimental | 0.5 | 1 | 0.5 | 2 |
| (M. Taoufik, 2015) | Journal | CAD, Neural Network, PCA, SVM | Mix Method | Observational | 0.5 | 0.5 | 0.5 | 1.5 |
| (P. Kathale and S. Thorat., 2020) | Conference | No | Qualitative | Ground Theory | 0 | 1 | 0 | 1 |
| (A.Rampun et al., 2019) | Conference | Contour-Based CNN approach | Quantitative | Experimental | 1 | 1 | 0.5 | 2.5 |
| (J.Wang et al,2016) | Journal | Deep Learning | Qualitative | Experimental | 1 | 1 | 0.5 | 2.5 |
| (Y. Zheng, 2020) | Journal | Computer aided detection using Gabor filtration | Qualitative | Experimental | 1 | 0.5 | 0.5 | 2 |
| (A. Jalalian et al., 2013) | Journal | Computer aided Detection | Qualitative | Observational | 0.5 | 1 | 0 | 1.5 |
| (Y. LeCun et al., 2015) | Journal | Deep learning | Quantitative | Observational | 0.5 | 0.5 | 0.5 | 1.5 |
| (J. Schmidhuber., 2015) | Journal | No | Qualitative | Ground Theory | 0 | 0.5 | 0 | 0.5 |
| (L. Deng and D. Yu, 2014) | Journal | No | Quantitative | Ground Theory | 0 | 1 | 0 | 1 |
| (A. Yadollahpour and H. Shoghi., 2015) | Journal | No | Quantitative | Observational | 0 | 0.5 | 0.5 | 1 |
| (D. Lévy and A. Jain, 2016) | Journal | Convolution neural networks | Qualitative | Experimental | 0.5 | 0.5 | 1 | 2 |
| (D. Guliato et al., 1998) | Conference | No | Quantitative | Observational | 0 | 1 | 1 | 2 |
| (S. AhmedMedjahed et al., 2019) | Journal | K-Nearest Neighbors algorithm | Qualitative | Experimental | 0.5 | 1 | 1 | 2.5 |

| Reference | Type | Method | Approach | Study Type | | | | |
|---|---|---|---|---|---|---|---|---|
| (M. Abed et al., 2018) | Journal | Muliti-fractal dimensions using Neural Network | Qualitative | Experimental | 1 | 1 | 1 | 3 |
| (J Thongkam et al,. 2008) | Journal | Random Forest | Qualitative | Experimental | 1 | 0.5 | 1 | 2.5 |
| (R lucht et al., 2002) | Journal | Segmentation based on Neural Networks | Qualitative | Experimental | 0.5 | 0.5 | 0.5 | 1.5 |
| (B zehng et al., 2014) | Journal | Hybrid K-SVM algorithm | Qualitative | Experimental | 1 | 1 | 1 | 3 |
| (V Ramen et al., 2010) | Journal [27] | Machine Learning | Quantitative | Observational | 0.5 | 1 | 0.5 | 2 |
| (H Jouni et al., 2016) | Journal [28] | Neural Network | Qualitative | Experimental | 0.5 | 0.5 | 0.5 | 1.5 |
| (I Muhic et al., 2013) | Journal [29] | Fuzzy c-mean analysis | Mix Method | Experimental | 1 | 1 | 1 | 3 |
| (B Mughal et al., 2018) | Journal [30] | NO | Quantitative | Ground Theory | 0 | 1 | 1 | 2 |
| (M. Alkhaleefah and C. C. Wu et al., 2017) | Journal [31] | Hybrid CNN and SVM approach | Qualitative | Experimental | 1 | 0.5 | 1 | 2.5 |
| (M. A. Al-Masni et al., 2017) | Conference [33] | Fully Connected Neural Network | Qualitative | Experimental | 1 | 1 | 0.5 | 2.5 |
| (D. A. Ragab et al.,2019) | Journal | Deep Neural Networks and support vector machines | Qualitative | Experimental | 1 | 1 | 1 | 3 |
| (X Zhang et al., 2018) | Journal 35 | DNN framework | Qualitative | Experimental | 1 | 1 | 0.5 | 2.5 |
| (J Arevalo et al., 2016) | Journal 36 | CNN using supervised learning | Qualitative | Experimental | 0.5 | 1 | 1 | 2.5 |
| (G. Carneiro, and A. P. Bradley., 2016) | Journal 37 | DNN Framework | Mix Method | Experimental | 1 | 1 | 1 | 2.5 |
| (M. A. Al-antari, et al., 2018) | Journal 38 | CNN model using VGG-C300 | Qualitative | Experimental | 1 | 1 | 1 | 3 |
| (P. Chatterjee et al., 2018) | Journal 39 | Bayesian Method and machine learning | Qualitative | Experimental | 1 | 0.5 | 0.5 | 2 |
| (S. Boudraa et al., 2020) | Journal 40 | CAD using super resolution reconstruction | Quantitative | Experimental | 1 | 1 | 0.5 | 2.5 |
| (A.H. Ahmad et al., 2018) | Conference 41 | D-CNN | Quantitative | Experimental | 1 | 1 | 0.5 | 2.5 |
| (W. E. Fathy and A. S. Ghoneim et al., 2019) | Journal 43 | DNN Framework | Mix Method | Experimental | 0.5 | 0.5 | 1 | 2 |
| (Bahiri et al., 2019) | Conference 44 | D-CNN, DesNet, SqueezeNet | Mix Method | Observational | 0.5 | 0.5 | 0.5 | 1.5 |
| (Y. I. A. Rejani and S. T. Selvi, 2009) | Journal | Support vector machines | Quantitative | Experimental | 0.5 | 1 | 1 | 2.5 |
| (S. Zhou et al, 2013) | Journal 47 | Shearlet feature extraction using SVM | Quantitative | Experimental | 1 | 0.5 | 1 | 2.5 |
| (M. Talha et al., 2019) | Journal 48 | Bayesian Classification | Mix Method | Experimental | 0.5 | 1 | 0.5 | 2 |
| (N. Safdarian and M. Hedyezadeh, 2019) | Journal 49 | Pattern recognition Technique | Qualitative | Experimental | 0.5 | 0.5 | 0.5 | 1.5 |
| (G. Murtaza et al., 2020) | Journal 50 | Deep learning | Mix method | Observational | 1 | 1 | 1 | 3 |
| (R. D. Ghongade and D. G. Wakde, 2017) | Conference 51 | Machine Learning using RF-ELM algorithm | Qualitative | Experimental | 1 | 1 | 0.5 | 2.5 |
| (A. R. Vaka., 2020) | Journal 53 | Machine learning model | Qualitative | Experimental | 0.5 | 0.5 | 0.5 | 1.5 |
| (V.K. Singh et al., 2020) | Journal | Generative adversarial CNN | Quantitative | Experimental | 1 | 0.5 | 0.5 | 2 |
| (G Danala et al., 2021) | Journal 56 | CAD using digital mammograms | Qualitative | Experimental | 1 | 0.5 | 1 | 2.5 |
| (T André and R. M. Rangayyan, 2006) | Journal 57 | Neural Network using edge sharpness and texture feature technique | Qualitative | Experimental | 1 | 1 | 1 | 3 |
| (V. Hariraj et al., 2018) | Journal 58 | SVM using fuzzy multilayer | Qualitative | Experimental | 1 | 1 | 0.5 | 2.5 |
| (A. Jouirou, 2019) | Journal | No | Quantitative | Observational | 0 | 1 | 0.5 | 1.5 |
| (F. A. Zeiser et al., 2020) | Journal 62 | Deep Learning and data augmentation | Mix Method | Experimental | 1 | 1 | 1 | 3 |
| (de Brito Silva et al., 2020) | Journal | No | Mix Method | Observational | 0 | 1 | 0 | 1 |
| (S. Saknure and D. Deshpande, 2020) | Journal 65 | Deep learning | Quantitative | Observational | 0.5 | 0.5 | 0.5 | 1.5 |
| (E. Elelimy and A. A. Mohamed, 2019) | Conference 76 | SVM, Gabor filer and SVD | Mix method | Observational | 1 | 0.5 | 0.5 | 2 |

## REFERENCES

[1] J. Dheeba, N. Albert Singh, and S. Tamil Selvi, "Computer-aided detection of breast cancer on mammograms: A swarm intelligence optimized wavelet neural network approach," J. Biomed. Inform., vol. 49, pp. 45–52, 2014.

[2] M. H. Yap et al., "Automated Breast Ultrasound Lesions Detection Using Convolutional Neural Networks," IEEE J. Biomed. Heal. Informatics, vol. 22, no. 4, pp. 1218–1226, 2018.

[3] "WHO | Breast cancer: prevention and control." [Online]. Available:https://www.who.int/cancer/detection/breastcancer/en/index1.html. [Accessed: 31-Jul-2020].

[4] P. Giri and K. Saravanakumar, "Oriental Journal of Breast Cancer Detection using Image Processing Techniques," Orient. J. Comput. Sci. Technol., vol. 10, no. 2, pp. 391–399, 2017.

[5] M. Taoufik, "Cad-Based Automated Carcinoma Detection and Classification in Breast," no. August, 2015.

[6] P. Kathale and S. Thorat, "Breast Cancer Detection and Classification," in International Conference on Emerging Trends in Information Technology and Engineering, ic-ETITE 2020, 2020.

[7] A. Rampun et al., "Breast pectoral muscle segmentation in mammograms using a modified holistically-nested edge detection network," Med. Image Anal., vol. 57, pp. 1–17, Oct. 2019.

[8] J. Wang, X. Yang, H. Cai, W. Tan, C. Jin, and L. Li, "Discrimination of Breast Cancer with Microcalcifications on Mammography by Deep Learning," Sci. Rep., vol. 6, no. 1, pp. 1–9, Jun. 2016

[9] Y. Zheng, "Breast Cancer Detection with Gabor Features from Digital Mammograms," Algorithms, vol.3, no. 1, pp. 44–62, Jan. 2010

[10] A. Jalalian, S. B. T. Mashohor, H. R. Mahmud, M. I. B. Saripan, A. R. B. Ramli, and B. Karasfi, "Computer-aided detection/diagnosis of breast cancer in mammography and ultrasound: A review," Clinical Imaging, vol. 37, no. 3. Elsevier, pp. 420–426, 01-May-2013

[11] Y. LeCun, Y. Bengio, and G. Hinton, "Deep learning," Nature, vol. 521, no. 7553, pp. 436–444, May 2015.

[12] J. Schmidhuber, "Deep learning in neural networks: An overview," Neural Networks, vol. 61, pp. 85–117, Jan. 2015.

[13] "What is Deep Learning and How Does it Work?" [Online]. Available: https://searchenterpriseai.techtarget.com/definition/deep-learning-deep-neural-network. [Accessed: 30-Jul-2020].

[14] L. Deng and D. Yu, "Deep Learning: Methods and Applications," Found. Trends® Signal Process., vol.7, no. 3–4, pp. 197–387, 2014.

[15] A. Yadollahpour and H. Shoghi, "Early Breast Cancer Detection using Mammogram Images: A Review of Image Processing Techniques Medical physicist at Bio-electromagnetic Clinic View project Disease progression modeling View project," Artic. Biomed. Pharmacol. J., 2015.

[16] "Various Types of Convolutional Neural Network | by Himadri Sankar Chatterjee | Towards Data Science." [Online]. Available: https://towardsdatascience.com/various-types-of-convolutional-neural- network-8b00c9a08a1b. [Accessed: 31-Jul-2020].

[17] "Convolutional Neural Networks Explained - Magoosh Data Science Blog." [Online]. Available:https://magoosh.com/data-science/convolutional-neural-networks-explained/. [Accessed: 31-Jul-2020].

[18] M. H. Yap et al., "Automated Breast Ultrasound Lesions Detection Using Convolutional Neural Networks," IEEE J. Biomed. Heal. Informatics, vol. 22, no. 4, pp. 1218–1226, Jul. 2018.

[20] D. Lévy and A. Jain, "Breast Mass Classification from Mammograms using Deep Convolutional Neural Networks," 2016.

[21] D. Guliato, R. M. Rangayyan, W. A. Carnielli, J. A. Zuffo, and J. E. L. Desautels, "Segmentation of breast tumors in mammograms by fuzzy region growing," vol. 20, no. 2, pp. 1002–1005, 2002.

[22] M. Abed, B. Al-khateeb, and A. Noori, "Neural network and multi-fractal dimension features for," Comput. Electr. Eng., vol. 0, pp. 1–12, 2018

[23] I. Muhic, "Fuzzy Analysis of Breast Cancer Disease using Fuzzy c-means and Pattern Recognition,"Southeast Eur. J. Soft Comput., vol. 2, no. 1, 2013.

[24] B. Mughal, M. Sharif, N. Muhammad, and T. Saba, "A novel classification scheme to decline the mortality rate among women due to breast tumor," Microsc. Res. Tech., vol. 81, no. 2, pp. 171–180, 2018.

[25] B. Zheng, S. W. Yoon, and S. S. Lam, "Breast cancer diagnosis based on feature extraction using a hybrid of K-means and support vector machine algorithms," Expert Syst. Appl., vol. 41, no. 4 PART 1, pp. 1476– 1482, 2014.

[26] V. Raman, "A Theoretical Methodology and Prototype Implementation for Detection Segmentation Classification of Digital Mammogram Tumor by Machine Learning and Problem Solving Approach," Int. J. Comput. Sci. Issues, vol. 7, no. 5, pp. 38–44, 2010.

[27] H. Jouni, M. Issa, A. Harb, G. Jacquemod, and Y. Leduc, "Neural Network architecture for breast cancer detection and classification," 2016 IEEE Int. Multidiscip. Conf. Eng. Technol. IMCET 2016, pp. 37–41, 2016.

[28] I. Muhic, "Fuzzy Analysis of Breast Cancer Disease using Fuzzy c-means and Pattern Recognition," Southeast Eur. J. Soft Comput., vol. 2, no. 1, 2013.

[29] [29]    B. Mughal, M. Sharif, N. Muhammad, and T. Saba, "A novel classification scheme to decline the mortality rate among women due to breast tumor," Microsc. Res. Tech., vol. 81, no. 2, pp. 171–180, 2018.

[30] M. Alkhaleefah and C. C. Wu, "A Hybrid CNN and RBF-Based SVM Approach for Breast Cancer Classification in Mammograms," Proc. - 2018 IEEE Int. Conf. Syst. Man, Cybern. SMC 2018, pp. 894–

[33] M. A. Al-Masni et al., "Detection and classification of the breast abnormalities in digital mammograms via regional Convolutional Neural Network," Proc. Annu. Int. Conf. IEEE Eng. Med. Biol. Soc. EMBS, pp. 1230–1233, 2017.

[34] D. A. Ragab, M. Sharkas, S. Marshall, and J. Ren, "Breast cancer detection using deep convolutional neural networks and support vector machines," PeerJ, vol. 2019, no. 1, p. e6201, Jan. 2019.

[35] X. Zhang et al., "Classification of whole mammogram and tomosynthesis images using deep convolutional neural networks," IEEE Trans. Nanobioscience, vol. 17, no. 3, pp. 237–242, 2018.

[36] J. Arevalo, F. A. González, R. Ramos-Pollán, J. L. Oliveira, and M. A. Guevara Lopez, "Representation learning for mammography mass lesion classification with convolutional neural networks," Comput. Methods Programs Biomed., vol. 127, pp. 248–257, 2016.

[37] N. D. B, G. Carneiro, and A. P. Bradley, "The Automated Learning of Deep Features for Breast Mass Classification from Mammograms," vol. 1, pp. 106–114, 2016.

[38] M. A. Al-antari, M. A. Al-masni, M. Choi, S. Han, and T. Kim, "International Journal of Medical Informatics A fully integrated computer-aided diagnosis system for digital X-ray mammograms via deep learning detection , segmentation , and classi fi cation," Int. J. Med. Inform., vol. 117, no. April, pp. 44– 54, 2018.

[39] P. Chatterjee, C. Mamatha, T. Jagadeeswari, and K. C. Shekhar, "Detection of Masses in Mammograms using Bayesian Method and Machine learning," vol. 7, pp. 108–111, 2018.

[40] S. Boudraa, A. Melouah, and H. F. Merouani, "Improving mass discrimination in mammogram-CAD system using texture information and super-resolution reconstruction," Evol. Syst., no. 0123456789, 2020.

[41] A. H. Ahmed and M. A. M. Salem, "Mammogram-Based Cancer Detection Using Deep Convolutional Neural Networks," in Proceedings - 2018 13th International Conference on Computer Engineering and Systems, ICCES 2018, 2019, pp. 694–699.

[42] N. Pradeep, "Segmentation and Feature Extraction of Tumors from Digital Mammograms," vol. 3, no. 4, 2012.

[43] W. E. Fathy and A. S. Ghoneim, "A deep learning approach for breast cancer mass detection," Int. J. Adv. Comput. Sci. Appl., vol. 10, no. 1, pp. 175–182, 2019.

[44] [A. P. Adedigba, S. A. Adeshinat, and A. M. Aibinu, "Deep learning-based mammogram classification using small dataset," in 2019 15th International Conference on Electronics, Computer and Computation, ICECCO 2019, 2019.

[45] Y. I. A. Rejani and S. T. Selvi, "Early Detection of Breast Cancer using SVM Classifier Technique," vol.1, no. 3, pp. 127–130, 2009.

[46] M. A. Al-masni et al., "PT US CR," Comput. Methods Programs Biomed., 2018.

[47] S. Zhou, J. Shi, J. Zhu, Y. Cai, and R. Wang, "Shearlet-based texture feature extraction for classification of breast tumor in ultrasound image," Biomed. Signal Process. Control, vol. 8, no. 6, pp. 688–696, 2013.

[48] M. Talha, G. Bin Sulong, and A. Alarifi, "Classification of breast mammograms into benign and malignant," Int. J. Multimed. Ubiquitous Eng., vol. 7, no. 2, pp. 359–363, 2012

[49] N. Safdarian and M. Hedyezadeh, "Detection and Classification of Breast Cancer in Mammography Images Using Pattern Recognition Methods," Multidiscip. Cancer Investig., vol. 3, no. 4, pp. 13–24, 2019.

[50] G. Murtaza et al., "Deep learning-based breast cancer classification through medical imaging modalities: state of the art and research challenges," Artif. Intell. Rev., vol. 53, no. 3, pp. 1655–1720, 2020.

[51] R. D. Ghongade and D. G. Wakde, "Detection and classification of breast cancer from digital mammograms using RF and RF-ELM algorithm," 2017 1st Int. Conf. Electron. Mater. Eng. Nano- Technology, IEMENTech 2017, 2017.

[52] W. Ma et al., "Breast Cancer Molecular Subtype Prediction by Mammographic Radiomic Features,"Acad. Radiol., vol. 26, no. 2, pp. 196–201, 2019.

[53] A. R. Vaka, B. Soni, and S. R. K., "Breast cancer detection by leveraging Machine Learning," ICT Express, no. xxxx, pp. 0–4, 2020.

[54] V. K. Singh et al., "Breast tumor segmentation and shape classification in mammograms using generative adversarial and convolutional neural network," Expert Syst. Appl., vol. 139, p. 112855, 2020.

[55] A. Mencattini, M. Salmeri, G. Rabottino, S. Member, S. Salicone, and S. Member, "Metrological Characterization of a CADx System for the Classification of Breast Masses in Mammograms," vol. 59, no. 11, pp. 2792–2799, 2010.

[56] G. Danala et al., "Classification of Breast Masses Using a Computer-Aided Diagnosis Scheme of Contrast Enhanced Digital Mammograms," Ann. Biomed. Eng., vol. 46, no. 9, pp. 1419–1431, 2018.

[57] T. C. S. S. André and R. M. Rangayyan, "Classification of breast masses in mammograms using neural networks with shape, edge sharpness, and texture features," J. Electron. Imaging, vol. 15, no. 1, p. 013019, 2006.

[58] V. Hariraj, W. Khairunizam, V. Vijean, and Z. Ibrahim, "FUZZY MULTI-LAYER SVM CLASSIFICATION," no. May 2019, 2018.

[59] T. C. S. S. André and R. M. Rangayyan, "Classification of tumors and masses in mammograms using neural networks with shape and texture features," Annu. Int. Conf. IEEE Eng. Med. Biol. - Proc., vol. 3, pp. 2261–2264, 2003.

[60] H. Cai et al., "Breast Microcalcification Diagnosis Using Deep Convolutional Neural Network from Digital Mammograms," Comput. Math. Methods Med., vol. 2019, 2019

[61] A. Jouirou, A. Baâzaoui, and W. Barhoumi, "Multi-view information fusion in mammograms: A comprehensive overview," Inf. Fusion, vol. 52, no. November 2018, pp. 308–321, 2019.

[62] F. A. Zeiser et al., "Segmentation of Masses on Mammograms Using Data Augmentation and Deep Learning," J. Digit. Imaging, 2020.

[63] T. F. de Brito Silva, A. C. de Paiva, A. C. Silva, G. Braz Júnior, and J. D. S. de Almeida, "Classification of breast masses in mammograms using geometric and topological feature maps and shape distribution," Res. Biomed. Eng., 2020.

[64] J. Dabass, M. Hanmandlu, and R. Vig, "Classification of digital mammograms using information set features and Hanman Transform based classifiers," Informatics Med. Unlocked, p. 100401, 2020.

[65] S. Saknure and D. Deshpande, "Multi-Scale Segmentation For Detecting Mass In Mammograms Using Deep Learning Techniques," SSRN Electron. J., no. Icicc, pp. 1–5, 2020.

[66] J. Deng, Y. Ma, D. ao Li, J. Zhao, Y. Liu, and H. Zhang, "Classification of breast density categories based on SE-Attention neural networks," Comput. Methods Programs Biomed., vol. 193, p. 105489, 2020.

[67] P. J. Sudharshan, C. Petitjean, F. Spanhol, L. Eduardo, L. Heutte, and P. Honeine, "Multiple instance learning for histopathological breast cancer image classification," Expert Syst. Appl., vol. 117, pp. 103– 111, 2018.

[68] A. Das, M. S. Nair, S. Member, and S. D. Peter, "Sparse Representation Over Learned Dictionaries on the Riemannian Manifold for Automated Grading of Nuclear Pleomorphism in Breast Cancer," IEEE Trans. Image Process., vol. 28, no. 3, pp. 1248–1260, 2019.

[71] A. Akselrod-ballin, L. Karlinsky, and S. Alpert, "A Region Based Convolutional Network for Tumor Detection and Classi fi cation in Breast Mammography," pp. 197–205, 2016.

[72] "CBIS-DDSM - The Cancer Imaging Archive (TCIA) Public Access - Cancer Imaging Archive Wiki." [Online]. Available: https://wiki.cancerimagingarchive.net/display/Public/CBIS-DDSM. [Accessed: 29- Jul-2020].

[73] R. S. Lee, F. Gimenez, A. Hoogi, K. K. Miyake, M. Gorovoy, and D. L. Rubin, "Data Descriptor: A curated mammography data set for use in computer-aided detection and diagnosis research," Sci. Data, vol. 4, Dec. 2017

[74] MIAS, "Mammographic Image Analysis Homepage - Databases." [Online]. Available: https://www.mammoimage.org/databases/. [Accessed: 29-Jul-2020].

[75] Attique, M., Farooq, M. S., Khelifi, A., & Abid, A. (2020). Prediction of therapeutic peptides using machine learning: computational models, datasets, and feature encodings. *Ieee Access*, 8, 148570-148594.

[76] E. Elelimy and A. A. Mohamed, "Towards Automatic Classification of Breast Cancer Histopathological Image," in Proceedings - 2018 13th International Conference on Computer Engineering and Systems, ICCES 2018, 2019, pp. 299–306.

[77] "Sci-Hub | Towards Automatic Classification of Breast Cancer Histopathological Image. 2018 13th International Conference on

Computer Engineering and Systems (ICCES) | 10.1109/ICCES.2018.8639219." [Online]. Available: https://sci-hub.tw/10.1109/ICCES.2018.8639219. [Accessed: 29-Jul-2020].

[78] S. K. Wajid and A. Hussain, "Local energy-based shape histogram feature extraction technique for breast cancer diagnosis," Expert Syst. Appl., vol. 42, no. 20, pp. 6990–6999, Nov. 2015.

[79] Mana R. Ehlers and Rebecca M. Todd, "Genesis and Maintenance of Attentional Biases: The Role of the Locus Coeruleus-Noradrenaline System," Neural Plast., vol. 1, no. 1, pp. 2–3, 2017.

[80] "(PDF) NOISE REMOVAL FROM DIGITAL MAMMOGRAM | EUROPEAN JOURNAL OF PHARMACEUTICAL AND MEDICAL RESEARCH." [Online]. Available:https://www.researchgate.net/publication/336771290_NOISE_REMOVAL_FROM_DIGITAL_MAMM OGRAM. [Accessed: 29-Jul-2020].